# Light scattering from exoplanet oceans and atmospheres

Short Title: Exoplanet light scattering

M. E. Zugger
Penn State University, Applied Research Laboratory
P.O. Box 30
State College, PA 16804
Center for Exoplanets and Habitable Worlds, The Pennsylvania State
University, University Park, PA 16802

J. F. Kasting
Penn State University, Department of Geosciences
443 Deike Bldg.
University Park, PA 16802
Center for Exoplanets and Habitable Worlds, The Pennsylvania State
University, University Park, PA 16802

D. M. Williams
School of Science; Penn State Erie, The Behrend College
4205 College Drive
Erie, PA 16563-0203
Center for Exoplanets and Habitable Worlds, The Pennsylvania State
University, University Park, PA 16802

T. J. Kane
Penn State University, Applied Research Laboratory
P.O. Box 30
State College, PA 16804
Penn State University, Department of Electrical Engineering
Electrical Engineering East
University Park, PA 16802

C. R. Philbrick
North Carolina State University
Physics Department, 432 Riddick Hall
Raleigh, NC 27695-8202

ABSTRACT

Orbital variation in reflected starlight from exoplanets could eventually be used to detect surface oceans. Exoplanets with rough surfaces, or dominated by atmospheric Rayleigh scattering, should reach peak brightness in full phase, orbital longitude = 180°, whereas ocean planets with transparent atmospheres should reach peak brightness in crescent phase near OL = 30°. Application of Fresnel theory to a planet with no atmosphere covered by a calm ocean predicts a peak polarization fraction of 1 at OL = 74°; however, our model shows that clouds, wind-driven waves, aerosols, absorption, and Rayleigh scattering in the atmosphere and within the water column, dilute the polarization fraction and shift the peak to other OLs. Observing at longer wavelengths reduces the obfuscation of the water polarization signature by Rayleigh scattering but does not mitigate the other effects. Planets with thick Rayleigh scattering atmospheres reach peak polarization near OL = 90°, but clouds and Lambertian surface scattering dilute and shift this peak to smaller OL. A shifted Rayleigh peak might be mistaken for a water signature unless data from multiple wavelength bands are available. Our calculations suggest that polarization alone may not positively identify the presence of an ocean under an Earth-like atmosphere; however polarization adds another dimension which can be used, in combination with unpolarized orbital light curves and contrast ratios, to detect extrasolar oceans, atmospheric water aerosols, and water clouds. Additionally, the presence and direction of the polarization vector could be used to determine planet association with the star, and constrain orbit inclination.

1. INTRODUCTION

1.1 The Search for Exoplanet Oceans

Since 1992 (Wolszczan & Frail 1992), astronomers have discovered over 480 planets[1] outside our Solar System (exoplanets), primarily through the radial velocity technique (which measures red and blue shifting of the starlight caused by motion of the parent star as the planet orbits around it). Once the candidate planets are confirmed, the Kepler mission[2] is expected to more than double this count by measuring small dips in stellar brightness due to planets passing in front of the star (the transit method). Recent papers have announced the direct imaging of multi-Jupiter-mass planets in orbits similar to or larger than that of Uranus (Kalas et al. 2008; Marois et al. 2008), but direct detection of Earth-sized planets in Earth-like orbits is not yet possible. However, space telescopes now under consideration, such as NASA's proposed Terrestrial Planet Finder—Coronagraph (TPF-C)[3], will eventually allow astronomers to capture reflected light from such planets and separate it from that of the parent star.

In addition to simply locating Earth-sized planets, we wish to determine whether they might be habitable and whether any of them might harbor life. Detection of liquid water on a planet's surface would be a strong marker of potential habitability. Although the presence of liquid water does not necessitate the presence of life, it is considered to be one of the best indicators of habitability because 1) liquid water requires both a significant atmosphere and moderate surface temperatures, and 2) the physical and chemical properties of liquid water make it an ideal medium for biochemical processes.

However, liquid water is not easily identified by remote detection. Various vibration-rotation bands of water *vapor* might be found in the near- and thermal-infrared of exoplanets (Des Marais et al. 2002; Tinetti et al. 2007), but exoplanet spectroscopy cannot easily distinguish between atmospheric and surface water. Long time series, multi-band photometry might potentially be used to identify land-ocean contrasts, as has been done for Earth using the EPOXI mission (Cowan et al. 2009); however, it is not clear that this technique will be feasible for the much fainter signals received from exoplanets. Here, we compute polarized and unpolarized light curves of water-covered planets, compare them with light curves expected from planets with Lambertian or dark surfaces, and determine the atmospheric and surface properties under which an ocean is detectable. We focus initially on idealized cases of planets with diffusely-reflecting Lambertian surfaces, dark surfaces, or specular-reflecting water surfaces, beneath varying cloud fractions and Rayleigh atmospheres of different optical thicknesses. We then calculate light curves of a few example water Earths which include Rayleigh scattering, aerosols, absorption, and waves.

---

[1] http://exoplanet.eu/catalog.php
[2] http://kepler.nasa.gov/
[3] http://planetquest.jpl.nasa.gov/TPF/tpf_index.cfm

1.2 Previous Modeling Efforts

Oakley & Cash (2009) modeled orbital and diurnal light curves of Earth-like exoplanets, but concentrated on planets with Earth-like geography, and did not study polarization. Mallama (2009) generated radiometric light curves for the terrestrial planets, but did not consider other types of planets or model polarization. Williams & Gaidos (2008) demonstrated that large oceans could be detected on exoplanets using the amplitude and shape of polarized and unpolarized orbital light curves. However, the model considered only surface scattering and did not include atmospheric effects. Also, the model assumed isotropic rather than Lambertian reflectance for diffuse scattering for clouds and rough surfaces. McCullough (2006) also modeled polarization, and included Rayleigh scattering, clouds, and different surfaces, but the work was unfortunately never published. Neither of the abovementioned polarization papers investigates the significant effects of absorption, aerosol scattering, scattering from within the water column, or varying degrees of ocean waviness, and neither paper compares the polarization signatures of ocean planets and dry planets. Dry planets with Rayleigh polarization signatures diluted by diffuse scattering might produce polarization signatures similar to those from ocean planets, resulting in false positives for the presence of oceans. Stam (2008) modeled Earth-like atmospheres over water surfaces, but he used a simple Fresnel model for oceans which does not include waves, sea foam, or scattering from within the water, and he did not model light curves of different atmospheres over an ocean. We seek here to extend the efforts of these previous workers by simulating polarized and unpolarized orbital lightcurves over a larger variety of atmospheric and ocean parameters.

## 2. MODEL DEVELOPMENT

### 2.1 Overview

We have coupled an atmospheric and surface modeling program (6SV)[4] with a planetary surface and orbital geometry program (Oceans). The 6SV atmospheric code was originally developed by the Laboratoire d'Optique Atmospherique and the European Centre for Medium Range Weather Forecast (Vermote et al. 2006), and was modified to compute polarization (Kotchenova & Vermote 2007; Kotchenova et al. 2006). The 6SV code calculates molecular scattering and absorption, aerosol scattering, and surface effects, including water surfaces with waves, and computes polarization effects in all of these scattering calculations. We have modified 6SV and written an IDL program which calls the modified 6SV and produces a lookup table of calculated scattering in both polarizations for thousands of combinations of stellar zenith angle, viewer zenith angle, and relative azimuth.

The Oceans code was developed originally by Williams, and used to simulate scattering from planets without atmospheres (Williams & Gaidos 2008). We have modified this code to use lookup tables generated by 6SV as inputs. The Oceans code computes the 3-D geometry of exoplanet orbits, calculates the light scattered to the observer at both polarizations from each 2° x 2° grid area on the planet, sums all of these contributions over the illuminated surface of the planet at each orbital point, and generates light curves and graphics illustrating orbit parameters, for planets without atmospheres. As modified for this work, the Oceans code also rotates the polarization reference plane from a ground-referenced system used by 6SV to the scattering-plane reference before summation.

### 2.2 6SV Model Details

The 6SV code is an atmospheric and surface radiative transfer code, written in Fortran, which simulates multiple scattering using successive orders of scattering. Coupling between the surface and the atmosphere is also included. The code approximates the vertical structure of the atmosphere with 30 layers, and solves the radiative transfer equation for each layer. An Earth-like pressure and temperature profile is assumed. The code performs numerical integration using decomposition in Fourier series for the azimuth angle, and using Gaussian quadratures for the zenith angle. The 6SV code uses 48 Gaussian angles to simulate scattering without aerosols; for aerosol simulations, 83 angles are used, including 0°, 90°, and 180°. Polarization is incorporated by calculating the first three Stokes parameters (I, Q, and U); circular polarization (V) is ignored. The effects of wavelength are included by using 20 node wavelengths between 0.25 and 4.0 μm and interpolating between them. We include absorption in the water Earth models; this is computed by 6SV for $O_3$, $H_2O$, $O_2$, $CO_2$, $CH_4$, and $N_2O$ using statistical band models with a resolution of 10 cm$^{-1}$. Earth-like altitude profiles are used for ozone and water vapor, and the other species are assumed to be well mixed.

---

[4] 6SV stands for Second Simulation of a Satellite Signal in the Solar Spectrum - Vector

We also include the effects of maritime aerosols in the water Earth models. Maritime aerosols are modeled after those found over Earth's oceans, and consist of a mixture of water droplets and crystals of sea salts. The parameters of maritime aerosols are as described in Levoni et al. (1997), and an exponential aerosol profile with a scale height of 2 km is assumed. The size distribution of the aerosols is then calculated by assuming a log-normal distribution, normalized so that the extinction coefficient at 550 nm corresponds to the visibility selected by the user; we use the standard visibilities of 5 km and 23 km, along with 80 km. The 6SV code then computes the aerosol scattering using the Lorenz-Mie solution.

2.3 Modifications to 6SV

The 6SV code has been used to calibrate the MODIS instruments on the Earth-observing satellites Aqua and Terra, and it was recently verified against other codes and actual data for some cases (Kotchenova & Vermote 2007; Kotchenova, et al. 2006). It has some inherent limitations, however, so we modified it as follows:

1) The code as written reports "apparent reflectance," which assumes a diffuse surface, and for ocean surfaces calculates a reflectance that increases without limit for large zenith angles. We modified the code so that it reports reflective Stokes parameters for ocean and diffuse Lambertian surfaces as values between zero and one, with a perfect mirror being assigned the value of unity. For ocean surfaces, we did this by replacing the wave tilt probability formula used in two subroutines[5] of 6SV, and adding appropriate output statements to the main routine;

2) The above change in the ocean surface model also made the model insensitive to wind direction, which should be viewed as beneficial, as this will generally be unknown and variable over time and space for an exoplanet;

3) We modified the Rayleigh scattering algorithm to allow exact round-number values of optical depth to be used for figures.

We also developed an algorithm to use the Kasten & Young (1989) equation to partially compensate for the difference between the plane-parallel and spherical atmosphere assumptions, but the resulting maximum differences in the integrated light curves was less than 1%.

2.4 Model Wavelength Ranges

To model hypothetical planets we must of course make some assumptions; with regard to wavelength, we do this in three different ways. The three cases are:

1) Lambertian surfaces, Lambertian clouds, and dark surfaces are assumed to be gray, and thus these results are wavelength independent of wavelength given that the albedo modeled corresponds to the albedo of the surface or cloud in the wavelength band of interest;

---

[5] These modifications will be discussed in detail in an upcoming paper.

2) For the water Earth models, we assume an Earth-like Rayleigh scattering atmosphere, molecular absorption, and maritime aerosols, and specify the baseline TPF waveband of 500 – 1000 nm;

3) For other cases, we parameterize Rayleigh scattering by atmospheres of hypothetical planets by showing how the planetary light curve changes with the Rayleigh optical depth $\tau_R$. When we parameterize by $\tau_R$, each curve for a given $\tau_R$ can represent a range of combinations of wavelengths and atmospheric densities. In order to give a sense of scale, the captions for these figures include the equivalent wavelength for an Earth-like Rayleigh scattering atmosphere corresponding to each value of $\tau_R$.

The caption for each figure includes the relevant wavelength range for cases 2) and 3), and for case 1), states that the curves are wavelength independent.

## 3. GEOMETRY AND DEFINITIONS

### 3.1 Orbital Geometry

In keeping with our desire to present idealized examples, we consider exoplanets with a single surface type and atmosphere type, in circular, edge-on orbits. Circular orbits are assumed both for simplicity and because such systems are presumed to be more likely to have stable climates suitable for continuously liquid water and life. We also assume that planets have a horizontally homogeneous atmosphere and a single surface type. A planet with a homogeneous surface in an orbit that is face-on to the observer (inclination, $i = 0$) has little variation with orbital phase, and so is not of interest. The effects we seek are maximized for edge-on inclinations ($i = 90$), so this is the case we model. This restriction is not unduly limiting, because half of all exoplanets will have orbital inclinations in the range $60 < i < 120$ because of geometrical considerations (Williams & Gaidos 2008). We will also assume an Earth-size planet at 1 AU from a Sun-like star.

The angle of polarization is defined relative to a chosen reference plane. In this case, we use the scattering plane, which is defined by the parent star, the planet, and the observer. For our edge-on geometry, the scattering plane is identical to the orbital plane. In Figure 1, it is also the plane of the paper. Although it appears at first glance that the scattering plane depends on what point on the star a particular light ray originates, and from where on the planet it is scattered, these effects are entirely negligible. One can understand this by noting that the apparent diameter of the Sun as seen from Earth is only about 0.5°, and the size of the Earth as seen from the Sun is 1 percent of that, and of course both objects appear very much smaller from the distance of another star system.

### 3.2 Polarization Fraction

Light propagating from the planet to the observer can be divided into components in which the electric field is parallel or perpendicular to the scattering plane. The difference between the perpendicular and the parallel components, divided by the sum, is called the polarization fraction:

$$PF = \frac{F_\perp - F_\parallel}{F_\perp + F_\parallel}. \quad (1)$$

### 3.3 Orbital Longitude

Orbital longitude (OL) is defined such that OL = 0° at new phase, when the planet passes in front of the parent star as seen from Earth (transit), and full phase (OL = 180°) occurs when the planet passes behind the star and is fully illuminated. When the planet appears farthest from the parent star, orbital longitude is 90° or 270°, and the planet is said to be at quadrature. If the planet (or moon) is close enough to be resolved, as with Mercury and Venus, the planet will be in crescent phase in the "front" portion of the orbit between 270° and 90°, in first or last quarter at 90° and 270°, and in gibbous phase in the "back" portion of the orbit, between 90° and 270°. We take advantage of orbital symmetry to simplify our curves by including only waxing phases (OL 0° to 180°) in our calculated light curves.

3.4 Orbital Longitude Effects on Polarization

Three orbital longitudes are of particular interest: OL = 74°, where the peak polarization of a flat water surface occurs, OL = 90°, where the peak polarization of Rayleigh scattering occurs, and OL = 140°, where the rainbow peak for water aerosols occurs. Polarization from Rayleigh scattering peaks when the source, scattering volume, and observer form a 90° scattering angle, a fact which can be verified by observing a clear blue sky at varying angles to the Sun with polarized sunglasses. Physically, the parallel component is unable to propagate to the observer in this geometry because the electric field vector is pointing in the direction of propagation. Therefore, for Rayleigh scattering in a thin atmosphere over a dark planet surface, the polarization fraction approaches 100%. If the atmosphere is thick, multiple scattering occurs, and the polarization fraction decreases. However, Chandrasekhar & Elbert (1954) calculated a number of cases, and found that polarization from a Rayleigh scattering atmosphere can exceed 90% before being limited by multiple scattering. In our case, the polarization fraction can be limited either by multiple scattering, by dilution from unpolarized or partially polarized light from the planet's surface, or both.

Simplistically, polarization from a flat water surface peaks at OL = 74°, as indicated in Figure 1. Light reflecting off a flat air/water interface at the Brewster angle, which for water is 53.1° to the normal, will also be polarized nearly 100%. Therefore, when the planet is at OL ~180 – (2 x 53°) = 74°, the polarization fraction from a flat water ocean (neglecting sea foam and scattering within the water column) is greatest. It is the parallel component that is not well reflected because it is oriented in the direction of travel. (This description is approximate and based on a simple conceptual model, but provides a useful starting point.) Our model below includes the effects of waves, sea foam, scattering from within the ocean, multiple Rayleigh scattering, clouds, and 3D geometry, and we compare various ocean planets against Lambertian land surfaces under Rayleigh atmospheres.

Returning to Figure 1, we see a third polarization peak at the rainbow angle OL = 140°, as described by Bailey (2007)[6]. This peak results from light interacting with approximately spherical airborne water droplets, as in a rainbow; light striking along the side of a droplet is refracted as it enters the droplet, reflects off of the inside back surface of the droplet, and refracts again as it leaves the droplet on the other side. The resulting angle between the incoming and outgoing light rays is 40°, resulting in a peak near OL = 140°.

---

[6] Our results may at first appear to disagree with Bailey's and Stam's results, but this is caused by differences in angle conventions. For orbital position, Bailey and Stam use phase angle, which is defined as the angle between the incident and outgoing light rays, so 0° occurs when the planet is on the opposite side of the star, or full phase for edge-on orbits. Orbital longitude (which we use) is defined such that 0° occurs when the planet is on the near side of the star, or in new phase for edge-on orbits. Therefore phase angle = 180 – OL.

## 4. UNPOLARIZED LIGHT CURVES AND CONTRAST RATIOS

First, we consider how total radiometric flux varies with orbital longitude. Figure 2 shows normalized light curves from three end member cases: a planet with a calm ocean under a thin atmosphere, a planet with a Lambertian surface under a thin atmosphere, and a planet with a single-scattering Rayleigh atmosphere over a dark surface. (The Lambertian surface is a mathematical approximation of a diffuse scattering surface; it assumes that reflectance drops off with the cosine of the viewing angle, so that it appears equally "bright" from any viewing angle, where "brightness" is measured in watts per steradian per square meter of projected area.) The resulting light curves are distinctive, especially the ocean planet light curve; hence, for terrestrial planets similar to the end member cases, unpolarized light curves by themselves could be useful in characterizing a planetary surface. We now discuss each of the three curves, and why their shapes are different.

### 4.1 Light curve descriptions

1) The Lambertian light curve closely matches the analytical result (Russell 1916; Sobolev 1975) and varies smoothly in an S-curve between zero flux and full flux as phase varies from new to full. At quadrature, the flux is $1/\pi \cong 0.32$ of that at full phase. The Lambertian planet is faint at small orbital longitudes because of the small illuminated surface fraction and the cosine weighting of the reflected flux.

2) The ocean light curve shows the opposite behavior: it peaks at small orbital longitudes (near $30^o$) when the planet is in crescent phase. This is because reflection from (calm) water is largest (~100%) near grazing incidence, and smallest (~2%) at normal incidence. The calm ocean curve was generated assuming a light wind of 1.5 m/s, which roughens up the surface enough so that nearly every illuminated pixel reflects some light to the observer. At full phase, OL = 180°, the entire face of the planet is illuminated, but with little reflectance. As the planet moves from full phase through the gibbous phase toward quadrature, the illuminated fraction becomes smaller, but the loss is mostly compensated by increased reflectivity. At orbital longitudes below $90^o$, the reflectance increases rapidly, much faster than the loss of illuminated surface, as the planet goes into crescent phase. The flux peaks near OL = $30^o$, where the incidence and reflection angles for specular (mirror-like) reflection are 75°, and the reflectance has increased tenfold to 20%. At OL < $30^o$, the loss in illuminated surface area dominates and the flux falls towards zero.

3) At small orbital longitudes, the normalized Rayleigh flux is higher than the normalized Lambertian flux, because the pathlength available for Rayleigh scattering becomes larger with increasing stellar zenith angles through the atmosphere. Both the normalized Rayleigh and Lambertian planet fluxes grow at high orbital longitudes because more of the observable planet surface is illuminated. (The Rayleigh flux curve here assumes an Earth-like atmospheric pressure profile; the shape of the curve would change somewhat with a different atmospheric pressure profile, but should remain distinct from the light curve of an ocean planet with a thin atmosphere and no clouds.)

4.2 Viewing considerations

Although it appears from Figure 2 that discriminating between three types of surface scattering is straightforward, it may be difficult to do in practice for several reasons. First, the planet can only be viewed when it is at sufficient separation from the star to be outside the inner working angle of the coronagraphic telescope. Exactly when that occurs will vary from one target to the next, but we can conservatively assume that most target planets will be observable only in the OL = 45 - 135° window. Second, the contrast ratio between the planet and the star must be sufficient to be able to observe the planet. Dark planets may not be observable, even at quadrature. And, third, real planets are likely to represent a combination of our three end-member cases, and so the respective light curves must somehow be deconvolved. Here, we discuss the specific issue of contrast ratios.

4.3 Albedo and Contrast Ratios

It is first important to define what is meant by *albedo*. The *Bond* albedo, or planetary albedo, is the fraction of all electromagnetic energy from the parent star that is scattered back into space at all angles. This is the albedo commonly used in energy balance and greenhouse effect calculations. Also relevant here is the *geometric* albedo, which is the ratio between the light reflected by a planet and the light that would be reflected by a white Lambertian disk. For a Lambertian surface, with the planet at full phase (OL=180°), the ratio between the geometric and Bond albedos is 2/3. Real planets have smaller or larger ratios between geometric and Bond albedo depending on surface type. The Bond albedo of Earth is 0.29 - 0.31, but the geometric albedo is 0.367 (Seidelmann 1992). Mars is closer to being a Lambertian planet, with a Bond albedo of 0.25 and a geometric albedo of 0.15[7].

The contrast ratio between an exoplanet and its parent star is an important factor in determining the observability of an exoplanet, and thus helps motivate the design of a planet-finding mission like TPF-C. The TPF Science and Technology Study Definition Team[8] calculated the contrast ratio $C_0$ between an Earth-sized Lambertian planet at 1 AU and its parent solar-type star at quadrature (OL = 90°) as:

$$C_0(90) = \frac{2}{3} A_{Bond} \frac{1}{\pi} \left(\frac{r_{planet}}{a_{1AU}}\right)^2 = A_{geo} \frac{1}{\pi} \left(\frac{6371 \text{ km}}{149.6 \times 10^6 \text{ km}}\right)^2 = 1.154 \times 10^{-10}. \quad (2)$$

The above calculation also assumes an Earth-like Bond albedo of 0.3. Table 1 lists example planets with Lambertian and water surfaces, and the corresponding contrast ratios at quadrature and full phase, relative to $C_0(90)$. Also, we follow McCullough (2006) in comparing the ocean planet values to the analytical result for a spherical mirror with a reflectance of 0.02, which is the value for water at normal incidence.

---

[7] http://ssd.jpl.nasa.gov/?planet_phys_par
[8] http://planetquest.jpl.nasa.gov/TPF/STDT_Report_Final_Ex2FF86A.pdf

The first three rows in Table 1 show that both our Lambertian cloud and surface models match the analytical results closely. The bottom three rows show that the calm ocean model at full phase gives close agreement with the spherical mirror approximation. Note that, at full phase, the brightness of the ocean planet is only about 8% of the brightness of the Lambertian planet. The ocean planet is also much dimmer than Earth, because Earth's albedo is dominated by clouds, with smaller contributions from Rayleigh scattering and from continental surface scattering.

4.5 The Terrestrial Planets

Mallama (2009) derived light curves for Mercury, Venus, and Mars, compensated for distance, and compared them to light curves based on earthshine from the Moon. All four terrestrial planet light curves decrease nearly monotonically as the planet moves from full phase to new phase, with Venus appearing three to four times as bright as the others. All four curves resemble some combination of our Rayleigh and Lambertian curves, except that Mercury shows a marked increase in brightness near full phase, and Venus shows a small flux increase near OL = 10°.

4.6 Summary of Radiometric Results

In summary, the Lambertian, Rayleigh-dominated, and ocean planets have widely differing unpolarized light curves that appear readily distinguishable. However it must be remembered that the ocean planet would be comparatively very dim and that observing at OL ≤ 45° may not be possible for many systems. Additionally, a realistic ocean planet would require a background atmosphere to hold onto its water, and it would almost certainly have clouds. It is nonetheless useful to simulate such idealized planets so that we understand the end-member cases. Also, the decreasing brightness of the water planet in the range OL = 45° to 90° should be a useful diagnostic. Because ambiguities will likely remain in the interpretation of radiometric (unpolarized) exoplanet light curves, we now address these by considering the polarization effects of various end-member planet types.

## 5. POLARIZED LIGHT CURVES

Figure 3 shows polarized light curves for a dark planet ($A_{bond}= 0$) with a thick Rayleigh atmosphere ($\tau = 0.5$) over a dark surface. The disk-averaged polarization fraction peaks near quadrature (OL = 90°) because Rayleigh scattering polarizes starlight to the greatest degree when it is scattered at right angles toward the observer. Planets with other Rayleigh optical depths over dark surfaces would have similar light curves, with the peak polarization fraction limited by dilution from the surface in cases of lower optical depths, and limited by multiple scattering for higher optical depths. A potentially useful measure of the contribution of Rayleigh scattering to the reflected flux is the phase lag between peak polarization and quadrature; planets with thin atmospheres and weak Rayleigh signatures will show polarization peaks at orbital longitudes less than (or sometimes greater than) 90°.

In Figure 4, we compare the light curves for planets with different Rayleigh optical depths. The planet surfaces are uniform Lambertian scatterers with a surface Bond albedo of 0.1, similar to that of Mercury and the Moon. The light curve for a $\tau = 0.05$ Rayleigh atmosphere over a black surface is shown for comparison (dashed curve). Light becomes increasingly polarized by Rayleigh scattering as optical depth increases, and the polarization maximum occurs nearer to quadrature as the atmosphere thickens. The effect is similar to that seen in Stam (2008) Fig. 4 (right-hand panel), although in that case the atmospheric density was held constant while the surface albedo was varied. The reference case in our figure (dashed), an atmosphere over a dark surface, shows the effect of removing surface backscattering, which otherwise dilutes the polarization fraction.

On a real planet with an atmosphere, scattering by clouds is also likely to be important. In Figure 5, we have replaced a fraction of each pixel on the planet with clouds, which are assumed to reflect light without polarizing it, in a Lambertian pattern (see section 6.5 for a brief discussion of this approximation). For simplicity we assume that the atmosphere above the clouds is thin enough to be ignored. We varied both cloud fraction and albedo, and found through comparative analysis (not shown) that the two parameters have nearly identical effect. Lambertian clouds dilute the polarization fraction and shift it towards smaller OL, as would a reflective Lambertian surface. Some clouds composed of liquid water droplets exhibit the rainbow angle effect; a planet with primarily these types of clouds would show a second polarization peak near OL = 140°. This "cloudbow" feature was measured for some Earth clouds by the POLDER instrument during aircraft-based testing (Goloub et al. 1994), and was also predicted for exoplanet water clouds by Bailey (2007) (see section 6.5).

Similar calculations were performed for ocean planets with thin atmospheres (Figure 6). As reflection from water-covered surfaces is affected by wind-driven waves and sea foam, the strength of winds is an important modeling parameter. For light winds (1.5 m/s), the polarization fraction approaches the value calculated by Fresnel theory for reflection at a smooth air/water boundary (dashed line). Our model departs slightly from the Fresnel equations because of the effects of sea foam (dependent on wind speed) and Rayleigh scattering (dependent on wavelength) from within the ocean. At wavelengths beyond 900 nm with the wind speed 1.5 m/s or less, the model results approach the Fresnel solution. As wind speed (and waviness) increases, the polarization fraction decreases and the OL of maximum polarization shifts to smaller values (Figure 7)[9]. Figure 7 was generated using the TPF waveband 500 – 1000 nm; when the model is run for a wavelength of 1000 nm and a wind speed of 1.0 m/s, the peak polarization of 0.978 at 74° (not shown). Here, waviness is parameterized using algorithms from Cox & Munk (1954) and wind speeds in the range 1-14 m/sec at 1 atm pressure.

In light of the discovery of methane oceans on Titan, it is reasonable to consider whether a liquid water surface could be confused with a surface covered by another liquid. Liquid methane has a refractive index of about 1.286 over the wavelength band of interest, resulting in a Brewster angle of 52.1°, corresponding to OL = 76°. For a methane ocean planet with thin atmosphere and no clouds, our model yields a peak polarization fraction of 0.902 at OL = 72°, very close to that for water. However, knowledge of the star's luminosity and the planet's orbital parameters should allow astronomers to distinguish between these two different liquids which have boiling points that differ by over 260 K at 1 atm.

In Figure 8, we compare the results of our model for 10 m/s winds (from Figure 7) with the results from Williams & Gaidos (2008). The primary differences between the 2008 ocean model and the current one are that the current model uses a slightly different parameterization of wind-generated sea foam, assumes Lambertian rather than isotropic scattering from sea foam, and includes scattering from within the water column.

As expected, a Rayleigh scattering atmosphere over an ocean surface produces a polarization fraction curve which is intermediate between the ocean-only and Rayleigh-only cases (Figure 9). Our model predicts that Rayleigh scattering from an Earth-like atmosphere with a pressure of 1 atm has a peak polarization at an orbital longitude of 83°, closer to the Rayleigh peak at 90° than to the Fresnel peak at 74°. This result is for the nominal TPF-C spectral band of 500-1000 nm. The Fresnel result is shown again for reference.

We consider the effect of clouds on the polarization signature of a calm ocean in Figure 10. The dilution effect of Lambertian clouds depends on the product of cloud fraction and cloud albedo. As with wind speed, increasing cloud fraction or albedo causes the polarization peak to decrease in magnitude and shift to smaller orbital longitudes. Clouds with the rainbow angle effect included should show a second peak near OL = 140°, and the overall peak may be shifted slightly to higher OL.

The polarization fraction curves of three water Earth models (Figure 11) demonstrate the significant effect of maritime aerosols even using a relatively high visibility of 23 km. Without aerosols, the water Earth polarization peak occurs at OL = 83°, as in the 1 atm curve in Figure 9 (the curve in Figure 11 adds absorption). Aerosols dilute the polarization peak, and add a shoulder near OL = 140° caused by the rainbow angle peak (Bailey 2007). This feature also appears in the "cloudy" planet light curves of Stam (2008) Fig. 9 (upper right-hand panel).

---

[9] In order to remove a nonphysical shoulder feature caused by numerical limitations of the model, Figures 7, 8, 9, and 10 were smoothed for OL = 2° to 6°, a portion of the orbit that will not typically be observable.

Figure 12 summarizes our polarized model results. We graph the point of peak polarization fraction for each model, thereby reducing each model case to a single point on the graph. This allows us to see trends as each parameter is varied. We now discuss these results in more detail.

The ocean planet results cluster around the point in the top center of the graph labeled "Calm Ocean, $\tau_R = 0$, No clouds." For wavelengths > 900 nm, this point would fall near the Fresnel peak polarization fraction of 1.0 at OL = 74°, and the position we calculate for the TPF band is caused by dilution by Rayleigh scattering within the water. The blue curve trending to the right shows the effects of increasing optical depth of a Rayleigh scattering atmosphere (see Fig. 8). Also, starting from the Fresnel result, the green curve shows that increasing Lambertian clouds (either cloud fraction or cloud albedo) reduces the polarization fraction, and moves the peak to lower OL. The violet curve shows the similar effects of increasing wind speed. The Williams and Gaidos 2008 result for 10 m/s wind is shown as a W, and lies near the triangle designating the equivalent 10 m/s result for the current model. The primary differences between that result and the current model are that Williams and Gaidos 2008 used different models for wave tilt and sea foam, did not include scattering within the water, and they used isotropic rather than Lambertian scattering for diffuse surfaces such as sea foam and clouds. The Lambertian model used here is a more accurate approximation of scattering from these features.

The black curve on the right side of Figure 12 shows results for a Rayleigh scattering atmosphere over a dark surface, with constant optical depth and increasing clouds. This curve models a planet dominated by Rayleigh scattering, with varying cloud cover. The polarization fraction drops and the peak moves toward smaller orbital longitude as more of the planet surface becomes cloudy, and as the albedo of the Lambertian clouds increases. The effects of increasing cloud fraction and increasing cloud albedo (not shown) are almost indistinguishable. For this curve, the surface reflectance is held constant at 0.001 (near the lower limit of model stability), and the atmospheric Rayleigh scattering optical depth is held at 0.5; however the shape of this curve is not strongly dependent upon either $\tau$ or surface reflectance as long as the Rayleigh scattering dominates and multiple scattering is present.

The red curve in Figure 12 shows polarization fraction from a Lambertian surface with reflectance of 0.10 with varying levels of Rayleigh scattering. The Lambertian surface curve approaches that for a dark surface when the same Rayleigh optical depth of 0.5 is assumed, because this atmosphere is thick enough to dominate the Lambertian surface reflectance of 0.1. As the optical depth is decreased, the peak polarization fraction drops and moves to lower orbital longitudes, as for the other cases.

Variations of water Earths are shown in the gray curve on the right side of Figure 12. For these models, we also include US 1962 Standard Atmosphere absorption (Dubin et al. 1962), and maritime aerosols with the standard visibilities, 23 km high visibility and 5 km low visibility. The surface is an ocean with a light wind speed of 1.5 m/s. The polarization fraction for these cases peaks at 0.149 at OL = 101° for the 5 km case, and at 0.353 at OL = 97° for the 23 km case. With more and more transparent aerosols, the polarization peak for water Earth cases approaches the blue curve representing a Rayleigh-only atmosphere over an ocean surface. As mentioned earlier, the polarization fraction for an ocean surface hidden by a Rayleigh-only atmosphere can be increased by using only the longer wavelength portion of the TPF waveband, from 900-1000 nm. However, this does not work for our more complex water Earth; when the calculations for a water Earth with 23 km visibility were repeated over the 900-1000 nm sub-band, the peak polarization fraction was 0.302 at OL = 97° (down from 0.353 at the same OL for the 500-1000 nm band ). For comparison, the sulfuric acid clouds of Venus give our neighbor planet a more complex (but diluted) polarization fraction, with peaks of about 0.02 near OL = 25° and OL = 165°, and a negative peak of about -0.035 near OL = 60° (Hansen & Hovenier 1974). These weak features would likely be lost in the noise when observing an extrasolar Venus.

If we were to use the position of the polarization peak in Figure 12 to determine whether or not an exoplanet is watery or dry, we might say that if the polarization fraction falls toward the upper left, the planet has an ocean surface, and if it falls in the lower right near the water Earth cases, it has water aerosols, and therefore has at least some water. Conversely, if the planet's polarization peak falls near the Lambertian or dark planet lines, then it is probably dry. Also, the position of the polarization peak should be interpreted in concert with the shape of the unpolarized light curve relative to the end-member cases given in Figure 2, and the overall brightness of the planet in comparison to the examples in Table 1. On the other hand, the gray curve shows that the peak polarization for a water Earth planet with thin aerosols can fall on or near the Lambertian and dark planet curves, producing a false negative. As demonstrated by the green and violet curves, clouds and waves shift the peak polarization point down and left on the chart, so a water planet with a combination of aerosols, clouds, and waves could have a peak polarization point falling virtually anywhere on the chart below the green and gray curves, including near the Lambertian or dark curves.

The remaining figures serve to verify our model against analytical results. Figures 13 and 14 verify that our model Lambertian light curve matches the analytical results of Sobolev (1975), and that the modified 6SV results for a calm ocean match the Fresnel curves. Figure 15 shows the glint spot from a water planet with a thin atmosphere and light wind (equivalent to 1.5 m/s at 1 atm) at OL = 74°; Figure 15a shows the parallel component, which is diminished at and near the Brewster angle, and Figure 15b shows the perpendicular component. The color scales, which represent the amount of flux scattered to the observer from each pixel, are the same – the parallel component peak is reduced by a factor of 3 due to Brewster angle effects. The vertical extent of the violet region is approximately +/-43° latitude, and the horizontal extent is approximately 58° of longitude. Pixels are defined as the area lying between 2° lines of latitude and longitude, so pixels shrink (and flux per pixel decreases) toward the poles as the cosine of latitude.

# 6. DISCUSSION

## 6.1 Effect of Orbital Inclination

As discussed earlier, the results shown here are for a planet with a homogeneous surface in an edge-on orbit (i = 90°). For increasingly face-on orbits, the variation in both total flux and polarization fraction over the planet's orbit is expected to approach zero, as was shown explicitly in Fig. 6 of Stam (2008). On the other hand, slightly inclined orbits could increase the fraction of the orbit in which the planet-star distance exceeds the minimum value that can be observed.

## 6.2 Effect of High Winds

Our model is limited to waviness caused by sea-level wind speeds up to approximately 14 m/s, the highest wind speed investigated by Cox & Munk (1954) measured at 41' (12 m) above the sea surface. An ocean planet with no land to impede the winds might conceivably have high wind speeds and a foamy, bright Lambertian surface. Alternatively, such a planet could conceivably have sustained winds from the same direction, creating a reflectance pattern that is highly asymmetric in azimuth, resulting in a complicated light curve that may be asymmetric in orbital longitude. This source of confusion might be minimized by observing over all orbital longitudes.

## 6.3 Wavelength-Dependent Light Curves

Using the Deep Impact spacecraft to observe the Earth as if it were an exoplanet, the EPOXI team (Cowan, et al. 2009) obtained light curves of Earth over seven 100-nm-wide wavebands between 300-1000 nm. Principal component analysis showed that two "eigencolors" captured 98% of the diurnal color changes caused by Earth's rotation. These eigencolors are essentially spectra of filters which could be used to distinguish between land surfaces and water surfaces. The cutoffs are gradual, but the land filter passes the red and near-infrared wavelengths between about 700 and 1000 nm, while the water filter passes the green, blue, and violet wavelengths below about 550 nm. Using this method the team was able to map the longitudinal variation in land surface area, which peaks when Africa and Europe are being observed and which approaches zero in the mid-Pacific. The technique proved successful despite the confusion generated by 50% cloud cover. This method of detecting land-sea contrasts might also work for an exoplanet, provided that enough photons were available to resolve diurnal variations.

## 6.4 Rainbow Angle

For planets with liquid droplets in the atmosphere, Bailey (2007) shows that at the "rainbow angle," total flux is higher and the polarization fraction can be as high as 0.2. For water droplets, the rainbow angle occurs at about OL = 140°, and for methane droplets, at OL = 131°. His calculations show that the effect is consistent for particle sizes from 10-100 $\mu$m at a wavelength of 400 nm, and that it weakens for smaller particles. These results are scalable throughout the TPF-C wavelength band when the ratio of particle size to wavelength is held constant. The Bailey paper predicts this effect for water droplets in clouds, but our model shows that the effect is similar for water aerosols such as the maritime aerosols typically found over Earth's oceans. Bailey's model and the 6SV aerosol model both use Lorenz-Mie scattering, so our result is a confirmation both of Bailey's result and of the 6SV aerosol model.

The effect of the rainbow angle for an ocean planet is to add a third competing polarization peak near OL = 140° to those caused by the water surface (near OL = 74°) and by atmospheric Rayleigh scattering (near OL = 90°). In our cloud-free water Earth model, the rainbow peak from aerosols shifts the total polarization peak to about 99°. We model clouds as simple Lambertian reflectors. If our cloud model included Lorenz-Mie scattering and exhibited a rainbow angle enhancement, clouds would serve not only to dilute the Fresnel water polarization peak, but to further strengthen the competing rainbow polarization peak. Such a cloud model may be developed for future analysis, however, clouds, even considering only those on Earth, are varied and complex (see section 6.5).

6.5 Clouds

Since seven of the eight solar planets and many of their moons have clouds of some type, we can reasonably expect most exoplanets with atmospheres to also have clouds. Across the solar system the variety of cloud types is impressive: sulfuric acid clouds on Venus, ice clouds on Mars, ammonia clouds on Jupiter and Saturn, and methane clouds on Neptune and Uranus. Earth clouds alone include a wide variety of water clouds with varying droplet sizes and distributions, as well as ice clouds made of particles ranging from highly symmetric hexagonal crystals to irregular particles. Each type of cloud has different scattering properties, so the parameter space for scattering from cloudy planets is enormous.

As mentioned earlier, some liquid water clouds have share the aerosol property of producing a polarized "cloudbow" near OL = 140°, which in some water-rich planets could reinforce the polarization signal from water aerosols. Ice clouds have different properties; work by Takano & Liou (1989, 1995) found that hexagonal and irregular ice clouds can exhibit a wide range of both local scattering peaks and polarization peaks depending on the particle size and shape distribution.

Rather than attempt to parameterize this range of characteristics of hypothetical exoplanet clouds, we have chosen to model clouds as Lambertian, a common practice in remote sensing of Earth, as pointed out by Acaretta et al. (2004). The Lambertian approximation is reasonable (especially for a broadband unpolarized stellar input) because, depending upon the variety of types of clouds present, the particle size range, and the amount of multiple scattering, the total cloud integrated signal can be nearly Lambertian. We discuss the cloudbow effect as one possible deviation from our model.

6.6 Exo-Zodiacal Light

Zodiacal light in our solar system is caused by scattering of sunlight by dust particles concentrated in and near the plane of the ecliptic. First detected around β-Pictoris by Smith & Terrile (1984), exo-zodiacal light (exo-zodi) scattered by similar disks in exoplanet systems is now considered to be a significant concern for future exoplanet investigation, and will generally be partially polarized. Additional data from surveys is needed to determine the characteristics of exo-zodi in order to optimize both future planet finding telescopes and analysis software.

6.7 Using Polarization to Determine Association and Orbital Inclination

The presence of significant linear polarization in the light from a candidate planet could be used in showing that the object is indeed in orbit around the parent star, and is not a background object. Regardless of whether the polarization is caused by Rayleigh scattering, a liquid surface, or both, a reduction in the "parallel" polarization component – the component parallel to the plane defined by the parent star, the planet, and the observer – would indicate that the object is probably a planet that is being illuminated by the star it appears to orbit. To date, polarization from giant exoplanets has been difficult to detect in the combined light of the planet and star (Lucas et al. 2009), but our model predicts significant levels of polarization for many different types of terrestrial planets when the planets can be resolved from the parent star.

Likewise, polarization could assist in determining the inclination of the orbit of planets with surfaces and atmospheres that are approximately horizontally homogeneous. For edge-on orbits, the polarization fraction varies widely as the planet orbits, but the direction of the polarization vector does not; for face-on orbits, the polarization factor is constant but the direction rotates with the planet's orbital motion. A combination of these two parameters could be used to determine the most likely orbital inclination angle. This effect would complement data from radial velocity measurement and from the planet's position angle in an image, and thereby help to determine the orbital inclination of the planet being observed.

6.8 The Inverse Problem

The variety of parameters which dilute, shift, or compete with the Fresnel polarization peak suggests that solving the inverse problem – determining from observations whether or not a planet has a large ocean – could be subject to numerous false positives and false negatives. Likewise, the end-member unpolarized light curves are distinctive, but an ocean surface with clouds, Rayleigh scattering, and aerosols can easily be dominated by any of the three atmospheric effects, or a combination of these and absorption. Still, some nearby ocean planets, if they exist, may have thin atmospheres and light cloud cover; other wet planets may be dominated by aerosols or water clouds, and detection of these would also be an indicator of potential habitability, whether or not they hide an ocean below.

## 7. SUMMARY AND CONCLUSIONS

We have developed a model for simulating orbital light curves from exoplanets which includes atmospheres with Rayleigh scattering, absorption, and aerosols, Lambertian clouds, dark and Lambertian surfaces, and ocean surfaces with varying degrees of roughness. We used this model to generate light curves for various hypothetical exoplanets in edge-on orbits, verified our model against analytical results, and compared our results to those of Williams & Gaidos 2008. We generated unpolarized total flux light curves for three end member cases: a Lambertian planet, a pure ocean planet, and a pure Rayleigh scattering planet with a dark surface. We also generated polarization fraction light curves for planets with various combinations of surfaces, atmospheres, and cloud cover. This work adds to the prior investigations by including clouds, varying wind-driven waves, Rayleigh scattering atmospheres, and Earth-like absorption and aerosol scattering; we also compare results for ocean planets with Rayleigh dominated planets and planets with diffuse scattering surfaces.

We conclude that while polarization by planetary oceans might be remotely detected on exoplanets with thin atmospheres, an atmosphere as thick as Earth's is enough to almost completely hide the water polarization signature when averaged over the TPF-C wavelength range (500–1000 nm) even without considering aerosols and absorption. Ocean radiance in this wavelength band caused by scattering within the water column also dilutes the polarization peak, limiting it to a maximum of about 0.9. The Rayleigh effects might be mitigated by using only the longer wavelengths of the TPF-C band, taking advantage of the dependence of Rayleigh scattering on the inverse fourth power of wavelength. However, this would result in throwing away ½ to ¾ of the available flux in the detector range, requiring either a larger telescope or longer integration times, and it would do nothing to reduce the dilution of the polarization signal by other factors. Still, if multiple wavebands are available on TPF-C, as baselined, then comparing the results of different wavebands from an exoplanet observation, with the above in mind, may be useful. Shifting to a slightly longer waveband is another possibility, although the silicon bandgap limits silicon detector technology to not much longer than 1000 nm, and the unpolarized black body emission from Earth rises at longer wavelengths.

Clouds also have a strong effect in masking the ocean surface polarization—a result that is not surprising, considering that the reflectance of water at near normal incidence is only about 2% across the wavelength band of interest. Waviness has a similar effect to that of clouds, hindering the detection of a polarization signal from the ocean. Exo-zodiacal light will likely contribute an additional "noise" polarization signal. The net effect of clouds, aerosols, absorption, atmospheric and oceanic Rayleigh scattering, and waves may severely limit the percentage of ocean planets that would display a significant polarization signature, and may also generate a significant number of false positives on dry planets.

All of this suggests that polarization measurements by a TPF-C type telescope may not provide a positive detection of surface liquid water on exoplanets. On the other hand, the placement of the polarization peak in Figure 12 relative to the curves shown there may give a strong hint of what type of planetary surface and atmosphere is being observed, especially when used in combination with the shape of the radiometric light curve relative to the three cases in Figure 2, and the overall planet contrast ratio relative to the cases in Table 1.

ACKNOWLEDGEMENTS

The authors gratefully acknowledge the efforts of the following in funding this research: Carl Pilcher of the NASA Astrobiology Institute; Vikki Meadows of the University of Washington and the Virtual Planetary Laboratory; Nick Woolf of the University of Arizona; and the Penn State Astrobiology Research Center.  The authors also thank Eric Vermote of the University of Maryland and NASA/Goddard for providing startup information on the 6SV code, and James Jenkins of Universidad de Chile for helpful discussions on determining planet association using polarization. The Center for Exoplanets and Habitable Worlds is supported by the Pennsylvania State University, the Eberly College of Science, and the Pennsylvania Space Grant Consortium. This work was performed under the following NASA Contracts: NNA04CC06A, NNX08A018G, and 463006 UNIWASH VPL.

Takano, Y., & Liou, K. 1989, Journal of the Atmospheric Sciences, 46
---. 1995, Journal of the Atmospheric Sciences, 52
Tinetti, G., et al. 2007, Nature, 448, 169
Tousey, R. 1957, J. Opt. Soc. Am, 47, 261
Vermote, E. F., Tanré, D., Deuzé, J. L., Herman, M., Morcrette, J. J., & Kotchenova, S. Y. 2006, Second Simulation of a Satellite Signal in the Solar Spectrum-Vector (6SV)
Williams, D. M., & Gaidos, E. 2008, Icarus, 195, 927
Wolszczan, A., & Frail, D. A. 1992, Nature, 355, 145
22

FIGURE CAPTIONS

Figure 1. Geometry of the problem, including planetary phase angles and analytical polarization maxima. Actual maxima differ when the spherical geometry and other factors are included.

Figure 2. Normalized total light curves for end-member planets. The peak near OL = 30° is indicative of an ocean surface (although not necessarily water).

Figure 3. Polarized light curves (solid) and polarization fraction (dashed) for a planet with a thick Rayleigh scattering atmosphere ($\tau_R$ = 0.5) over a dark surface. The parallel component has a local minimum at 90°; polarization fraction is limited to about 0.7 at about OL = 90° primarily by multiple scattering. Parallel and perpendicular are defined relative to the scattering plane, which is also the plane of the orbit for our edge-on cases. [Equivalent wavelength for an Earth-like Rayleigh scattering atmosphere is approximately 369 nm.]

Figure 4. Polarization fraction variation with $\tau_R$ over 0.1 Lambertian surface; pure Rayleigh over dark surface shown (dashed) for comparison. Unpolarized light from the surface dilutes polarization caused by Rayleigh scattering, so atmospheres with larger $\tau_R$ have a higher polarization fraction. [Equivalent wavelengths for an Earth-like Rayleigh scattering atmosphere are approximately 369 nm, 547 nm, 648 nm, 813 nm, 965 nm, and 1150 nm.]

Figure 5. Effect of Lambertian cloud fraction on Rayleigh scattering-dominated planet; $\tau_R$ = 0.5 [Earth atmosphere at 369 nm], a relatively low average cloud albedo of 0.3 is assumed. Unpolarized light from clouds dilutes polarization from Rayleigh scattering. Some clouds with liquid water droplets may exhibit the rainbow angle effect as a second polarization peak near OL = 140°.

Figure 6. Polarized light curves (black) and polarization fraction (gray) for planet with calm ocean and thin atmosphere, wavelength range 500–1000 nm. The Fresnel curve for light reflecting from a flat air/water boundary is shown for comparison. The model departs from Fresnel theory primarily because of scattering from within the water at the shorter wavelengths.

Figure 7. Polarization fraction for ocean planets vs. wind speed (no absorption, no aerosols, no Rayleigh scattering). For the TPF-C wavelength range 500–1000 nm, at wind speeds below 5 m/s, polarization fraction is limited by scattering within the water; at higher wind speeds, sea foam is the primary limiting factor on polarization fraction.

Figure 8. Comparison of polarization fraction results from Williams & Gaidos (2008) Fig 7b and the current model with a thin atmosphere and waves equivalent to Earth oceans with a surface wind of 10 m/s . The primary differences between the ocean models are that the current model uses a slightly different parameterization of wind-generated sea foam, uses Lambertian rather than isotropic scattering from sea foam, and includes scattering from within the water column. [Current model run with wavelength range 500–1000 nm; Williams & Gaidos 2008 model is wavelength independent.]

Figure 9. Polarization ratio for ocean planets vs. Rayleigh optical depth in equivalent Earth atmospheres for the TPF waveband 500-1000 nm. Both atmospheric and in-water scattering are significant at the shorter end of this band. Wind is calm (1.5 m/s), and absorption and aerosols are neglected.

Figure 10. Polarization fraction for ocean planet with varying cloud fraction and cloud albedo. As with clouds and a Rayleigh atmosphere (Figure 5), unpolarized scattering from clouds dilutes polarization signatures. [Wavelength range 500–1000 nm.]

Figure 11. Water Earth models with varying aerosols. Aerosol versions show large reduction in the peak polarization, even for relatively clear 23 km visibility; also exhibit a shoulder near OL = 140° due to rainbow angle peak. All three curves assume an ocean planet with Earth atmosphere Rayleigh and absorption, and a surface wind speed of 1.5 m/s. [Wavelength range 500–1000 nm.]

Figure 12. Summary of polarization fraction results for various planet types over the TPF waveband (500-1000 nm). Green curve is ocean with increasing clouds; violet curve is ocean with increasing surface wind; blue curve is ocean with increasing Rayleigh atmospheric depth. Red curve is Lambertian surface with increasingly thick Rayleigh scattering atmosphere; black curve is a thick Rayleigh atmosphere over a dark surface, with increasing clouds. W is Williams & Gaidos 2008 result for an ocean planet with 10 m/s winds. Water Earths (gray curve) include 1.5 m/s wind, Earth Rayleigh scattering, and US 1962 absorption model.

Figure 13. Contrast ratio between a Lambertian planet with Bond albedo of 0.3 and the parent Sun-like star at 1 AU. Curves from our model and the analytical result from Sobolev (1975) and extended in the TPF report (footnote 7) are essentially identical. [wavelength independent]

Figure 14. Reflectance of a calm ocean surface (wind speed = 1.5 m/s) to the parallel and perpendicular components of incoming radiation in the range 900 – 1000 nm, and the resulting polarization fraction. Curves for the modified 6SV closely match the Fresnel curves; the slight deviation in the polarization fraction near the Brewster angle (53.1°) is caused by small amounts of sea foam and scattering within the water column, which dilute the polarization.

Figure 15. a) Relative flux per pixel, **parallel polarization**, produced at and near the glint spot for a water planet with a thin atmosphere and no clouds. Planet is located at OL = 74°.
b) Relative flux per pixel, **perpendicular polarization**, same scale as in a), shows factor of 3 increase in peak flux.

TABLE

**Table 1**
Contrast ratios for Lambertian and ocean planets. Rows in ***bold italics*** show results from analytical calculations for comparison.

| Surface | Atmosphere | C(90°) Rel. to Lamb. | C(180°) Rel. to 2% mirror | Comments |
|---|---|---|---|---|
| ***Lambertian, $A_{bond} = 0.3$*** | ***none*** | ***1.000*** | ***…*** | ***from Sobolev (1975)/ TPF report*** |
| Cloud Lamb., A = 0.3 | none | 0.997 | … | cloud albedo = 0.3, cloud fraction = 1 |
| Surface Lamb., $\rho = 0.3$ | none | 0.998 | … | surface $\rho$ = 0.3, cloud fraction = 0 |
| ***Spherical mirror, $\rho = 0.02$*** | ***none*** | ***…*** | ***1.00*** | ***from Tousey (1957)*** |
| ocean 1.5 m/s 900 - 1000 nm | none | 0.079 | 1.06 | Rayleigh reduced using longer wavelengths |
| ocean 1.5 m/s 500 – 1000 nm | none | 0.089 | 1.34 | Significant Rayleigh scattering within water |

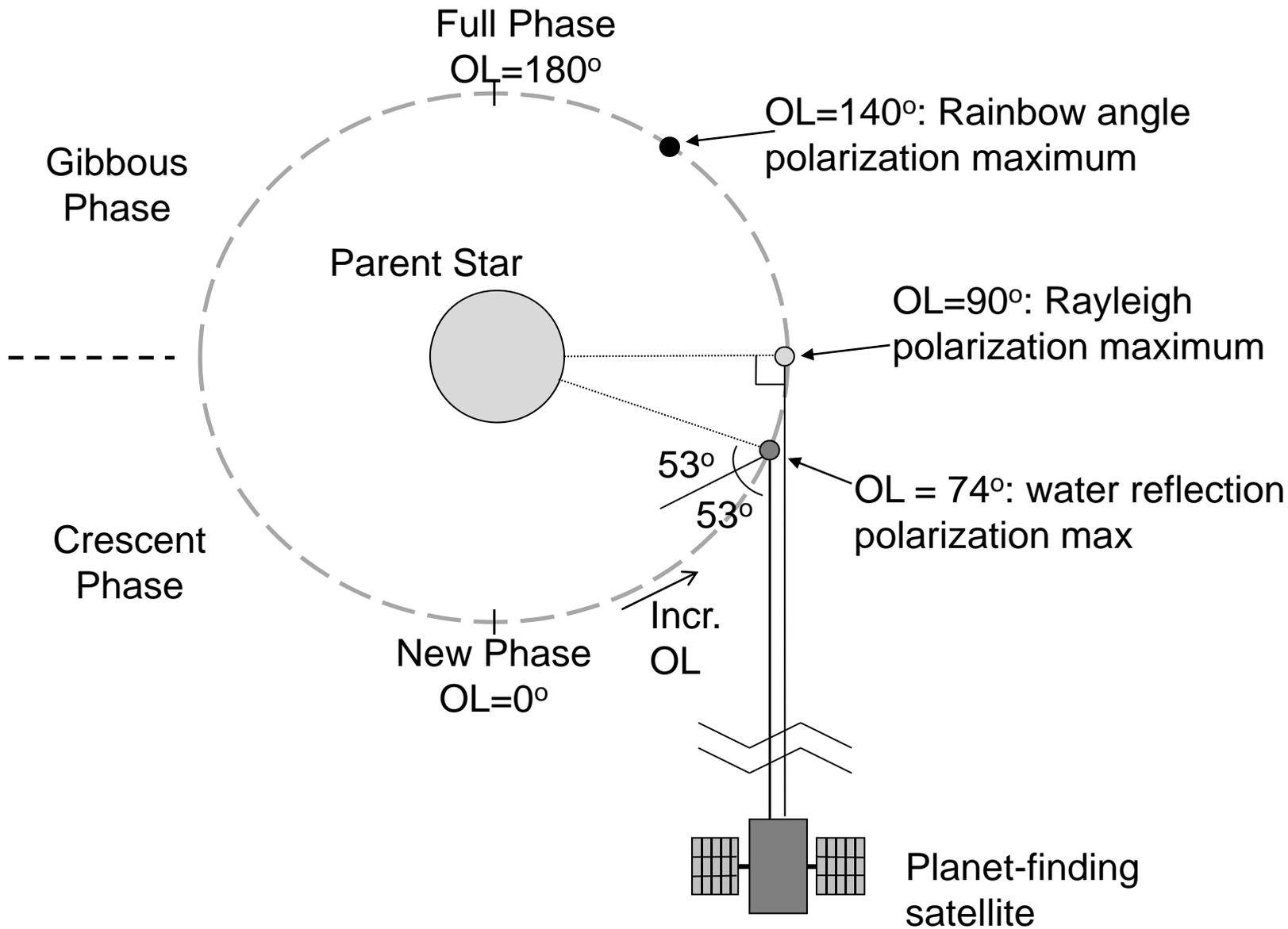

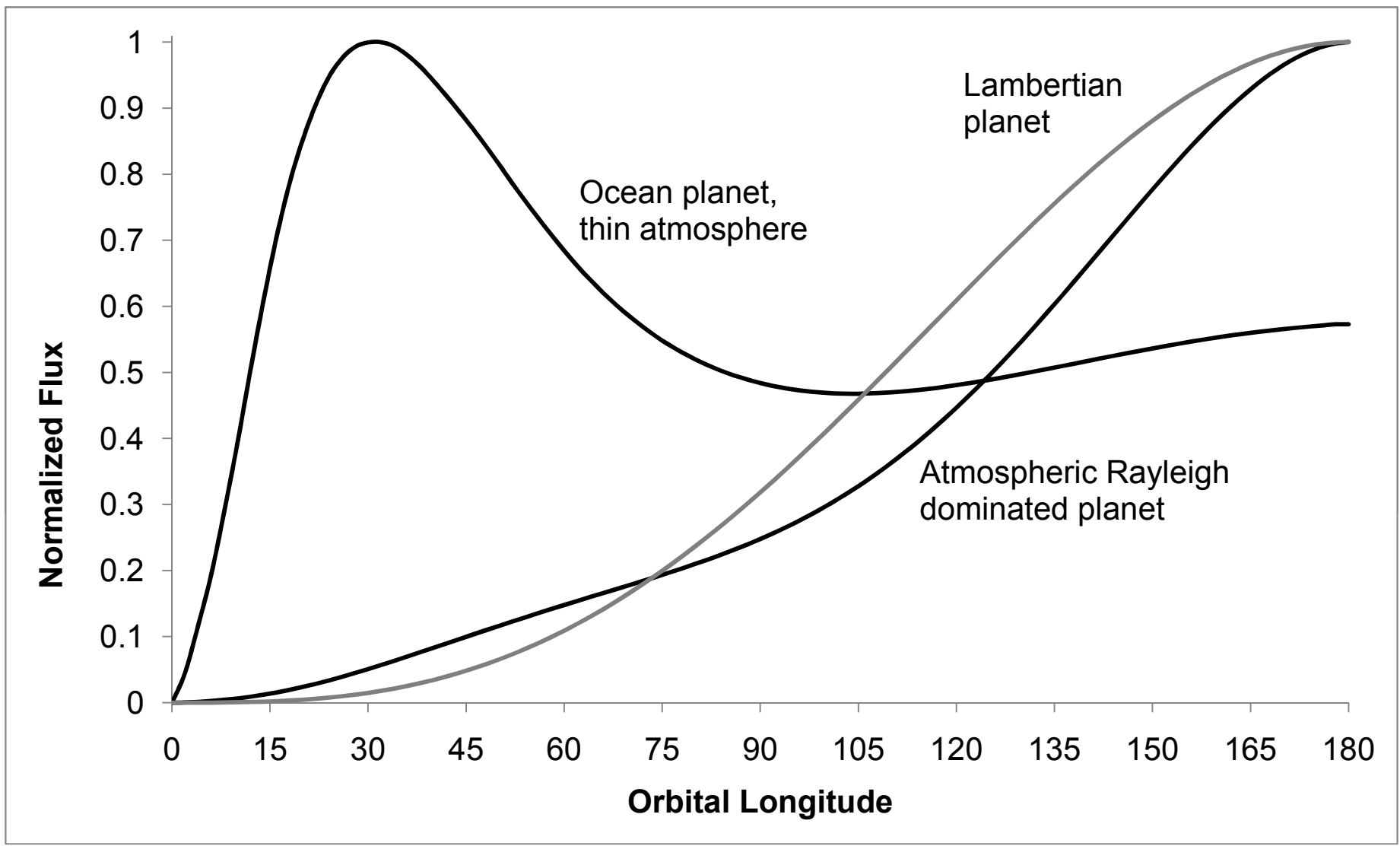

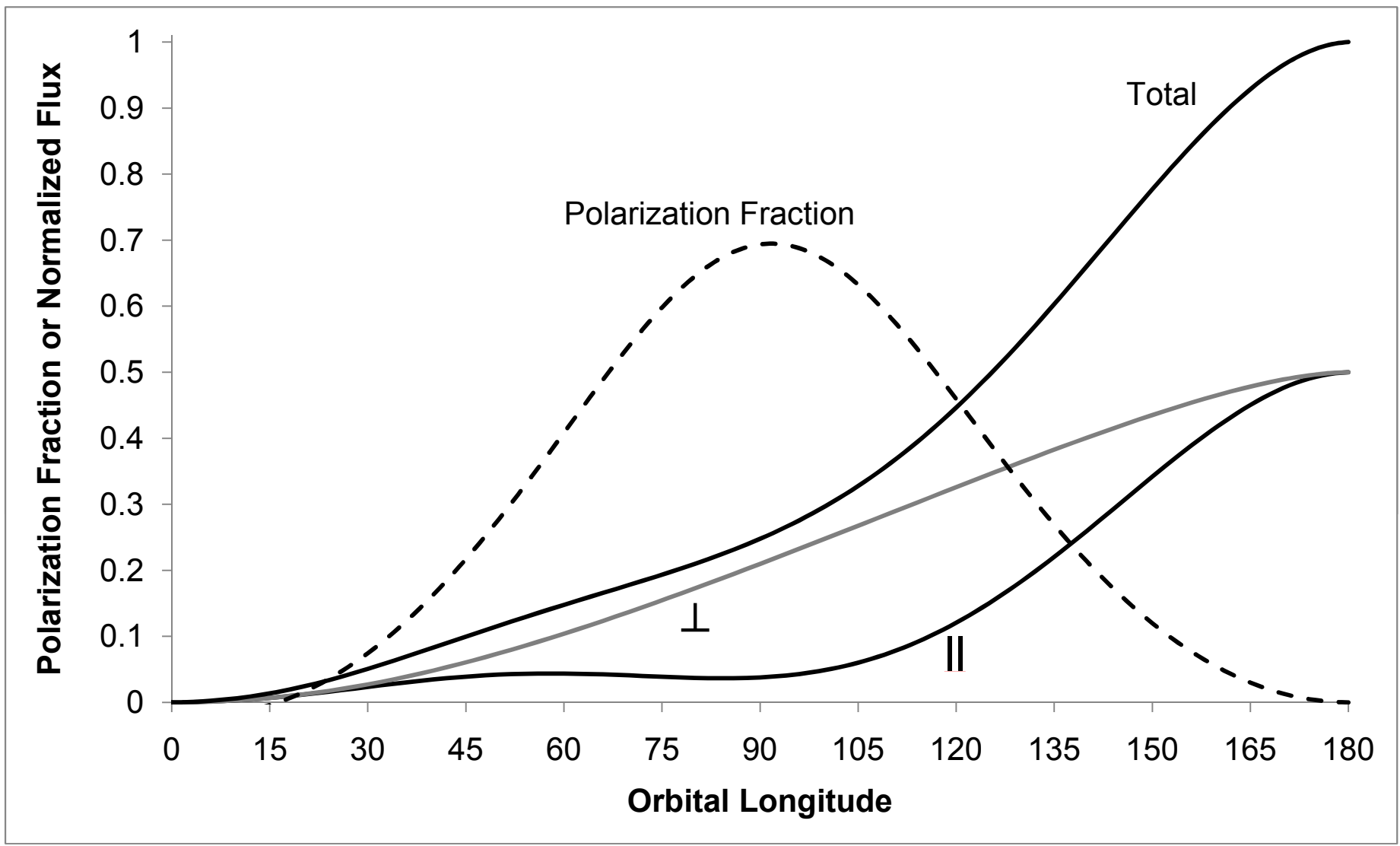

Polarization Fraction vs. Orbital Longitude

Rayleigh over dark surface, albedo ≈ 0, $\tau_R = 0.5$

Varying Rayleigh optical depth $\tau_R$ over Lambertian surface, albedo = 0.1

$\tau_R = 0.5$
0.1
0.05
0.02
0.01
0.005

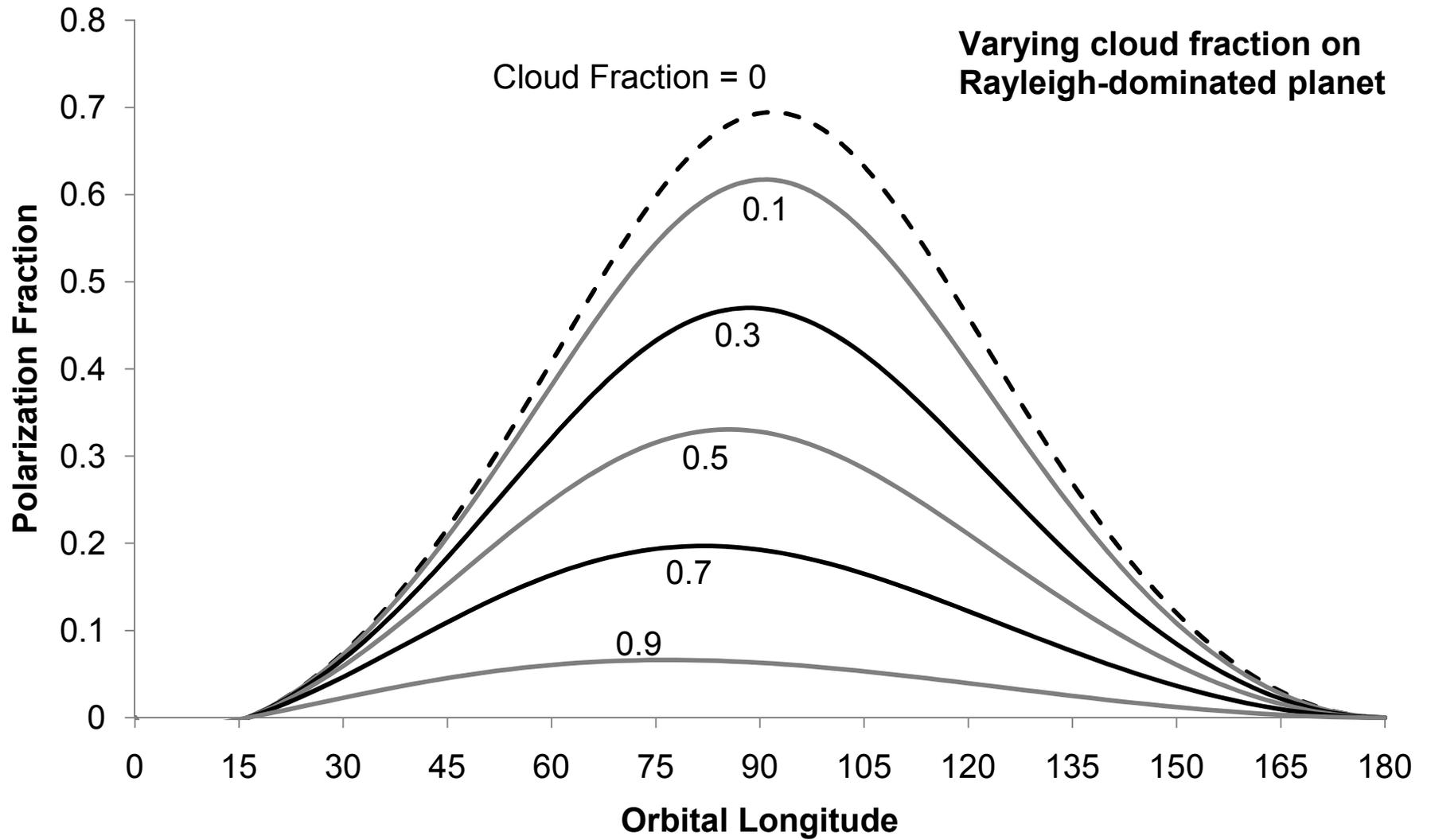

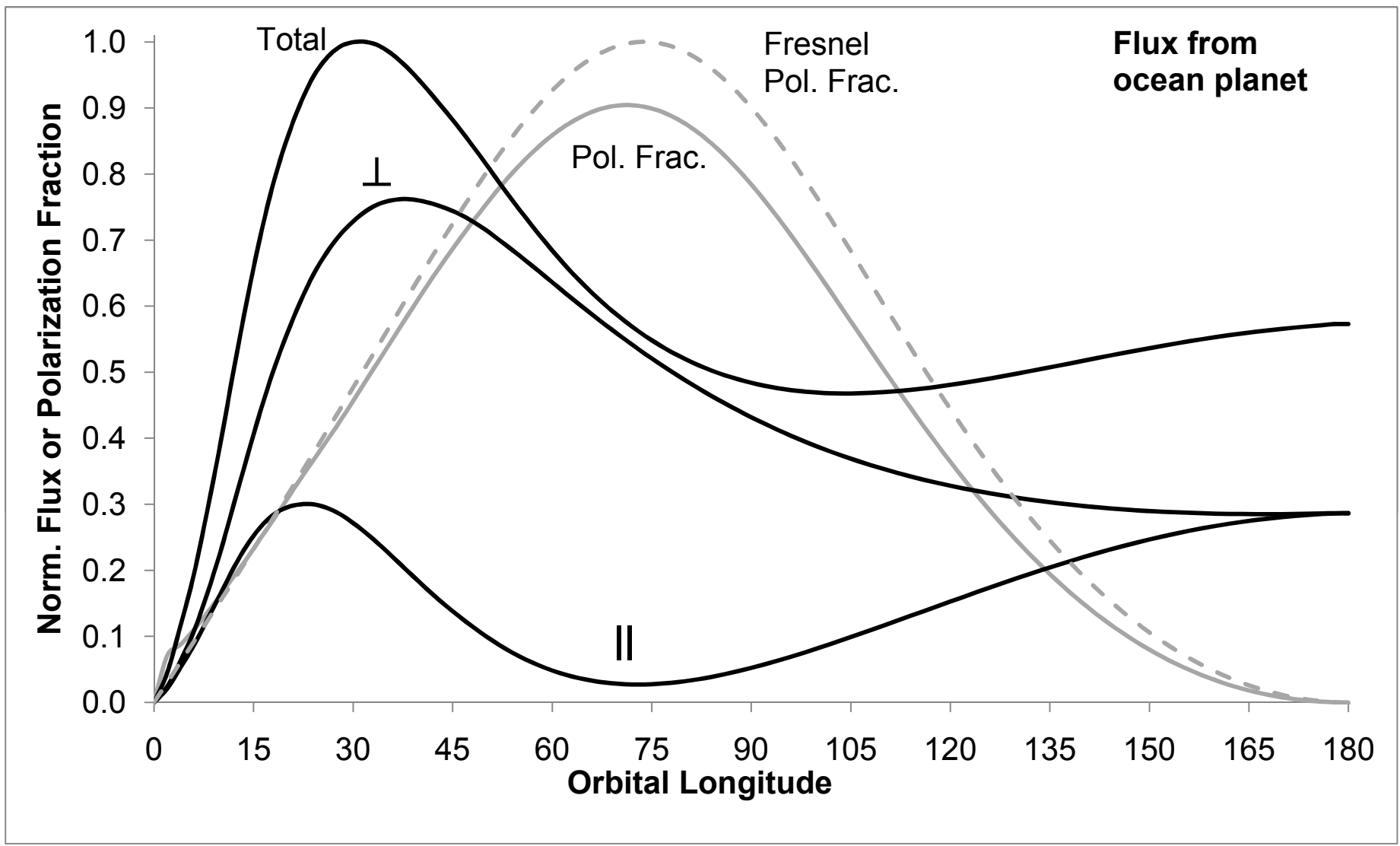

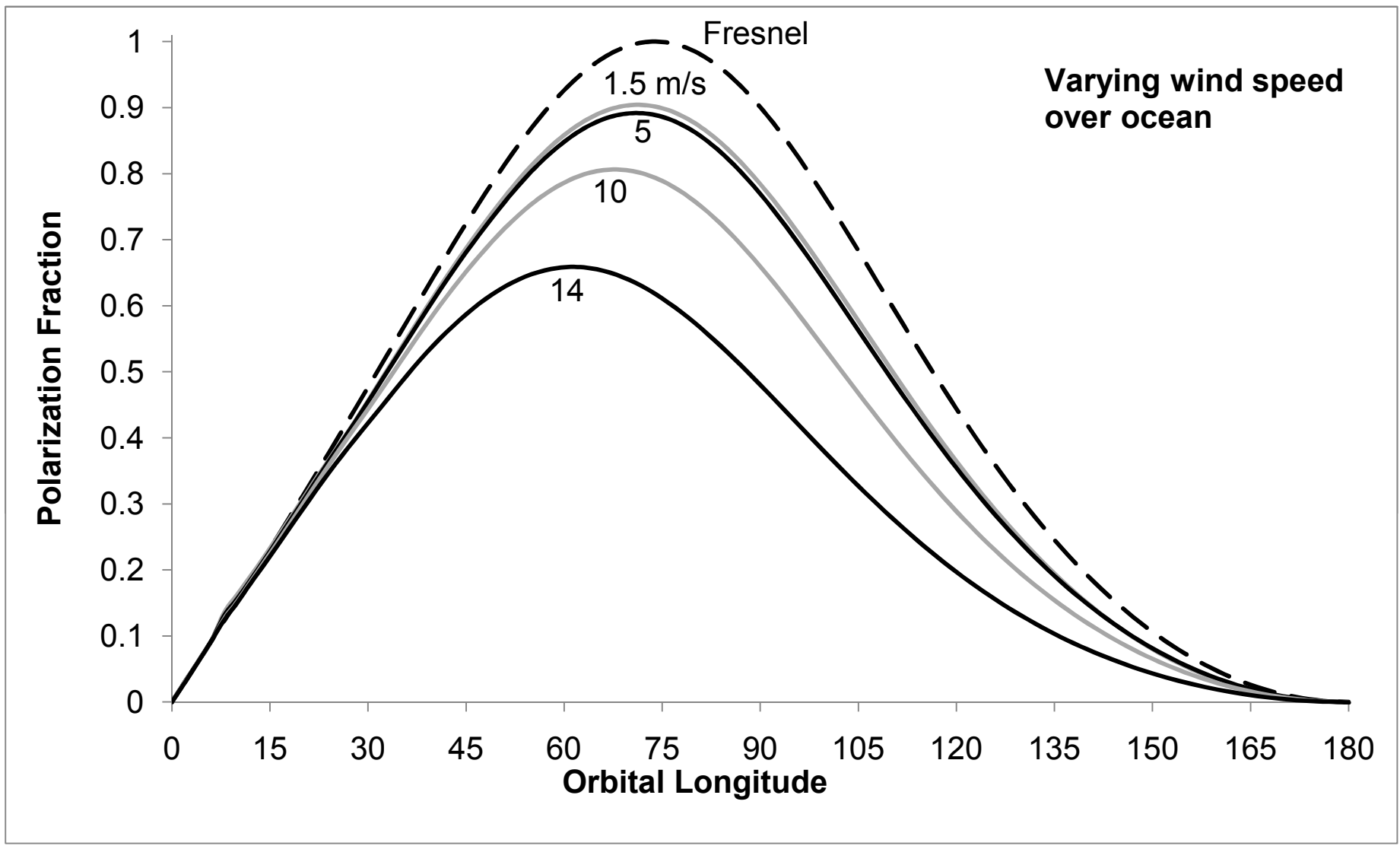

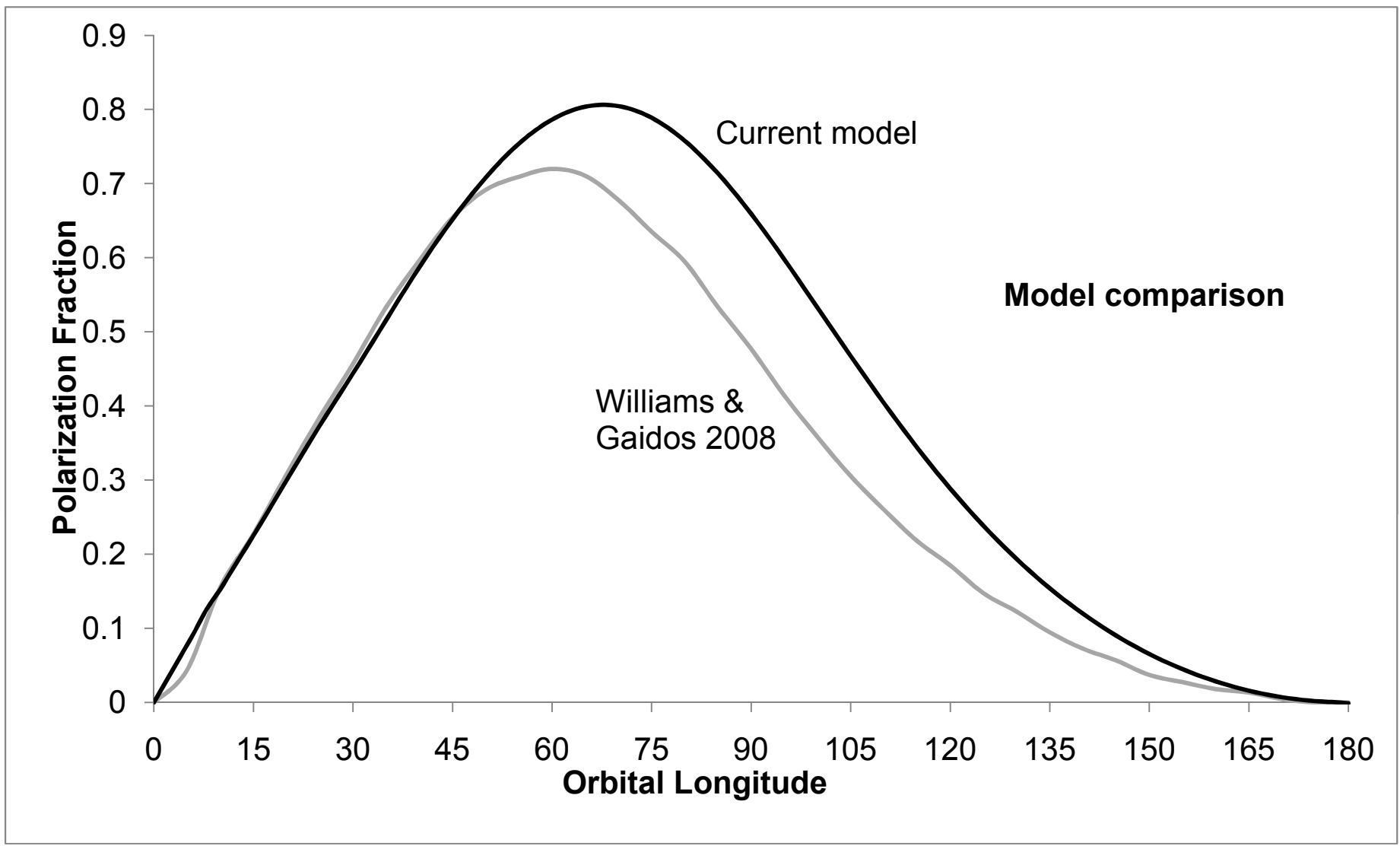

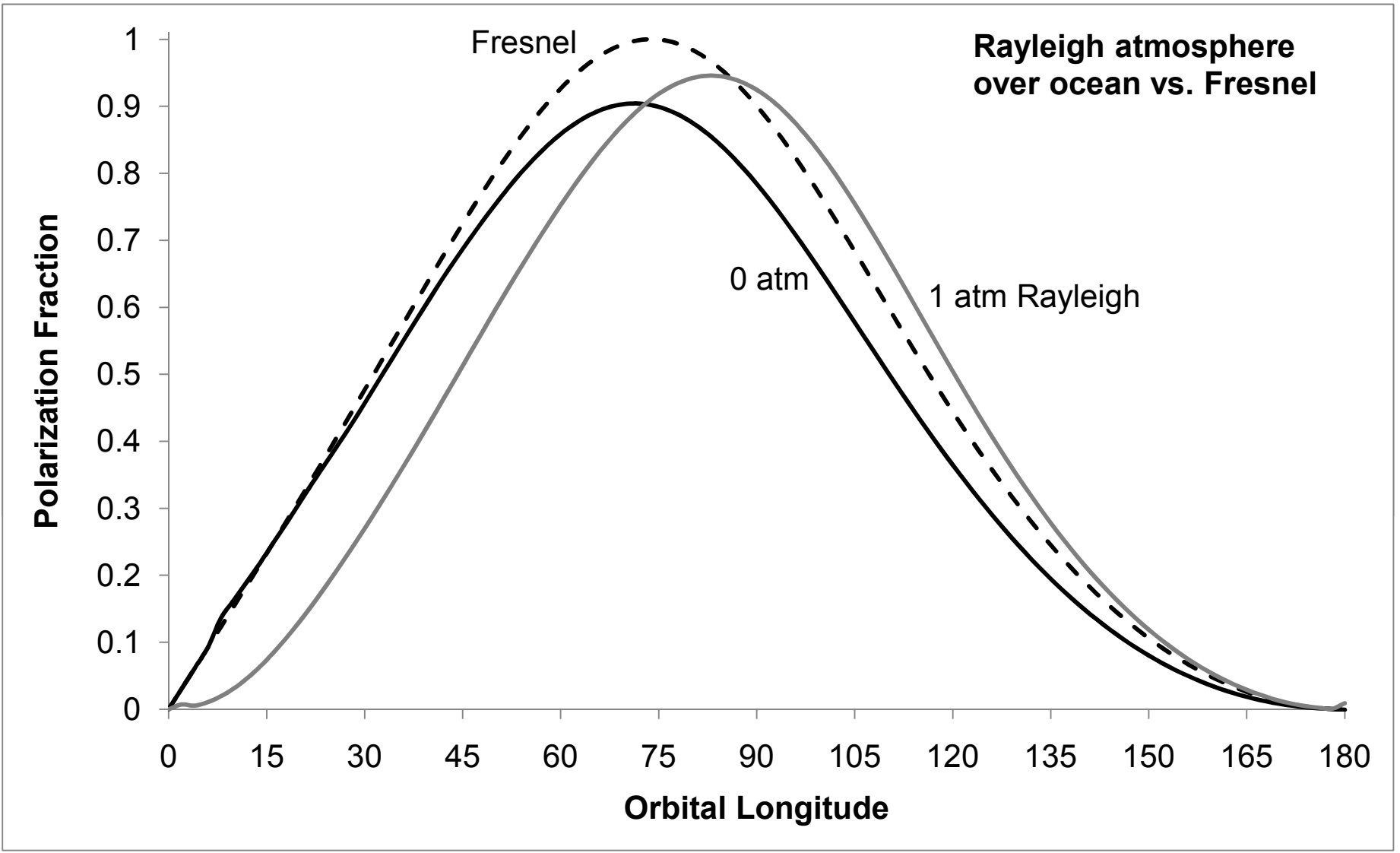

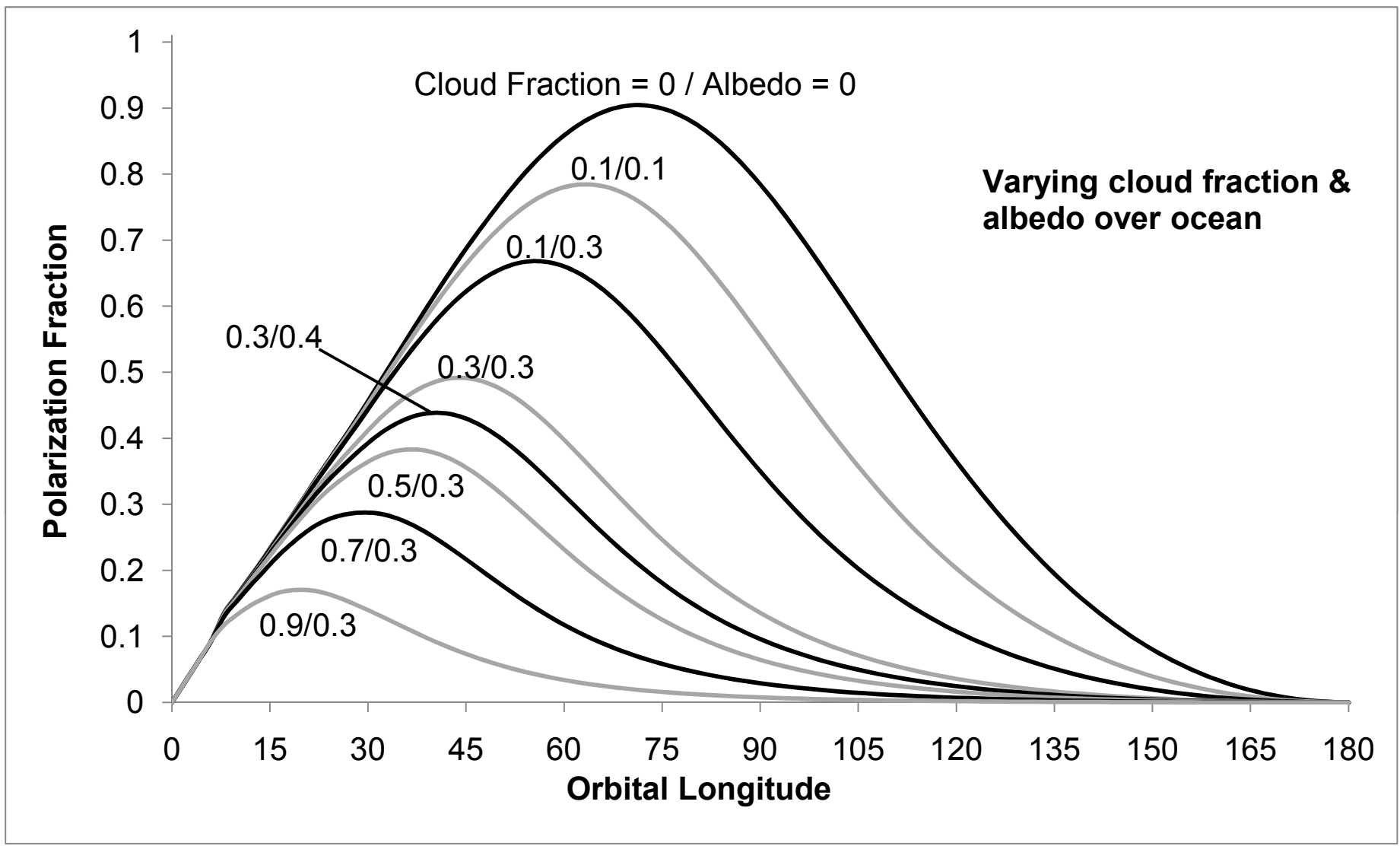

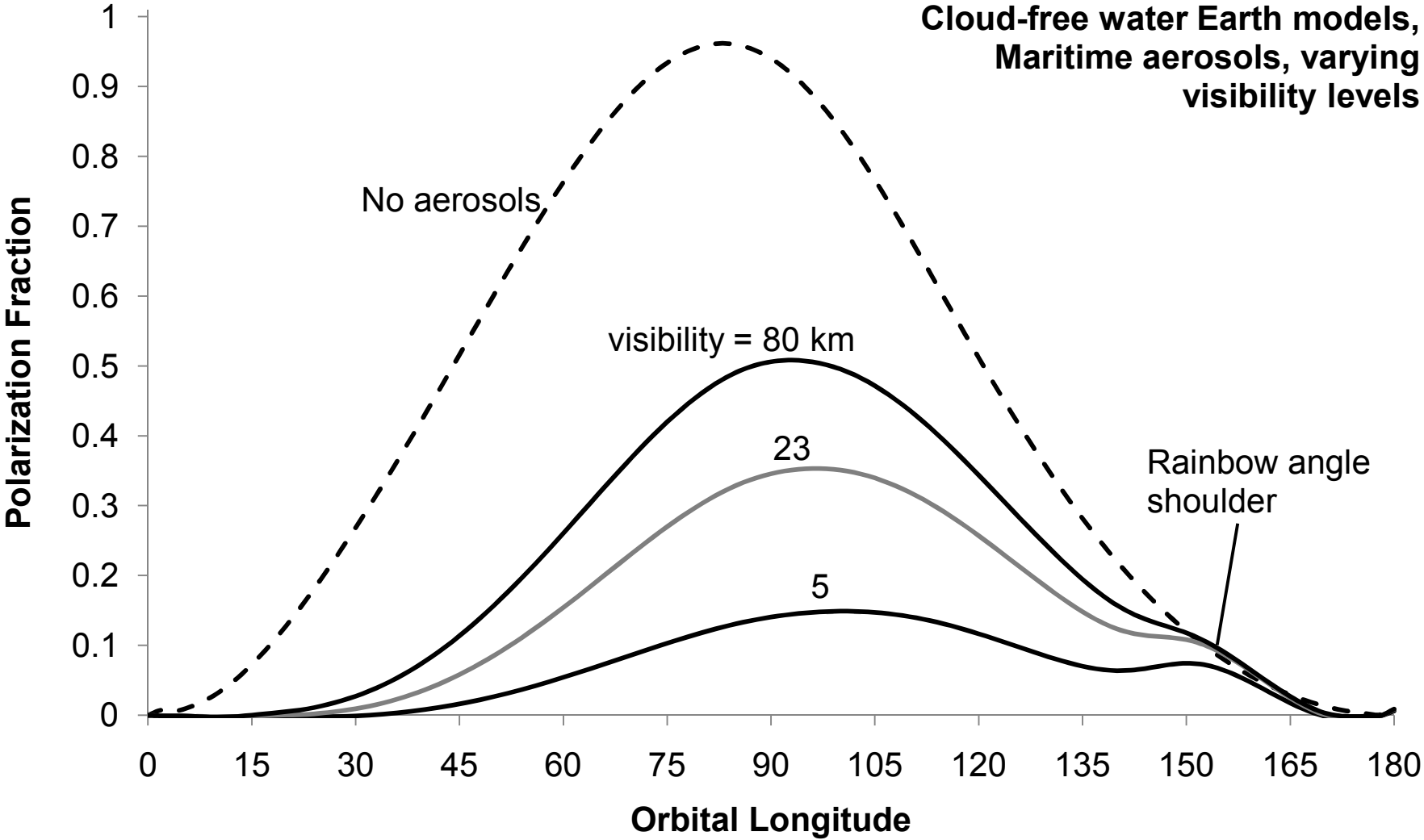

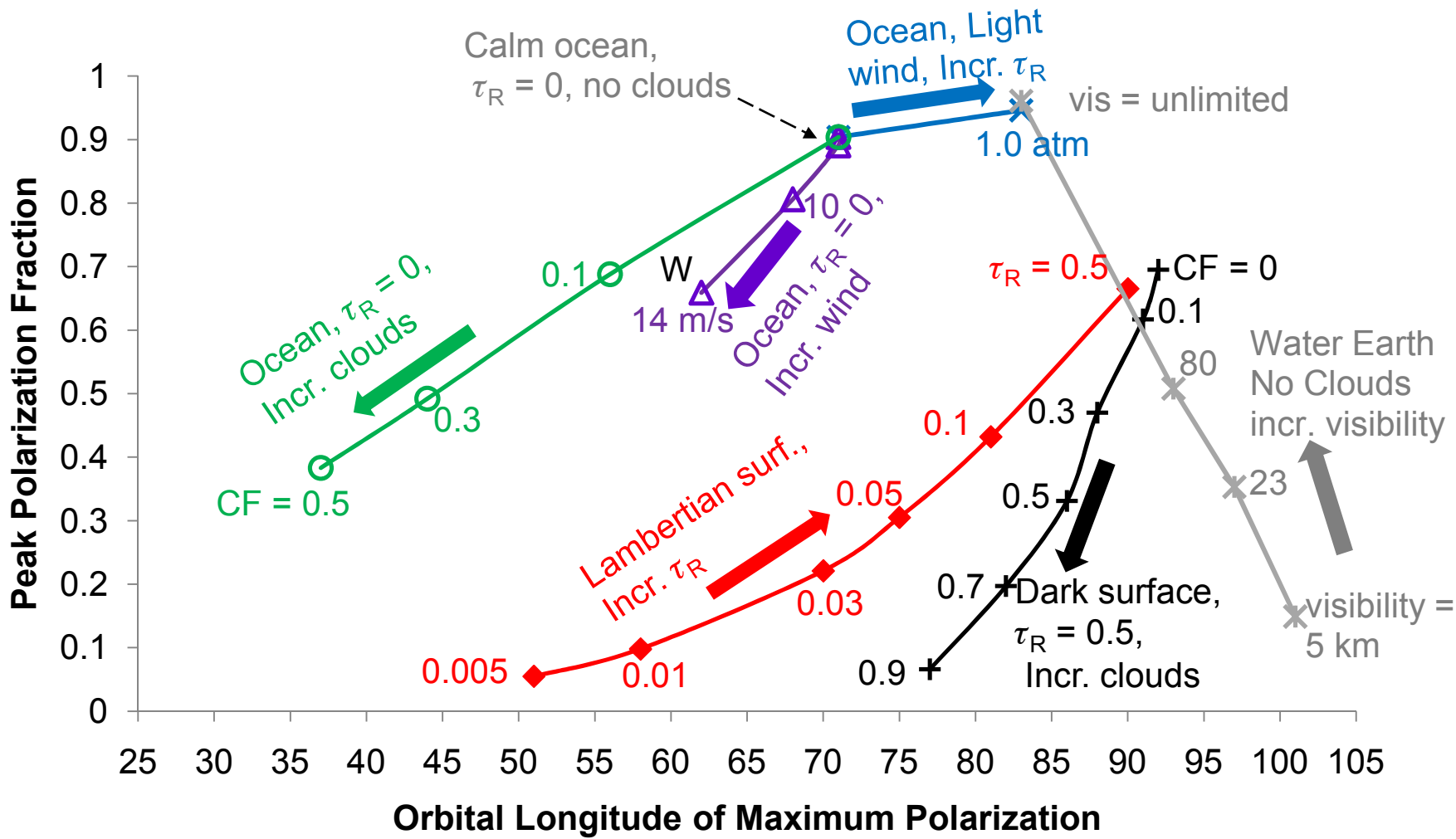

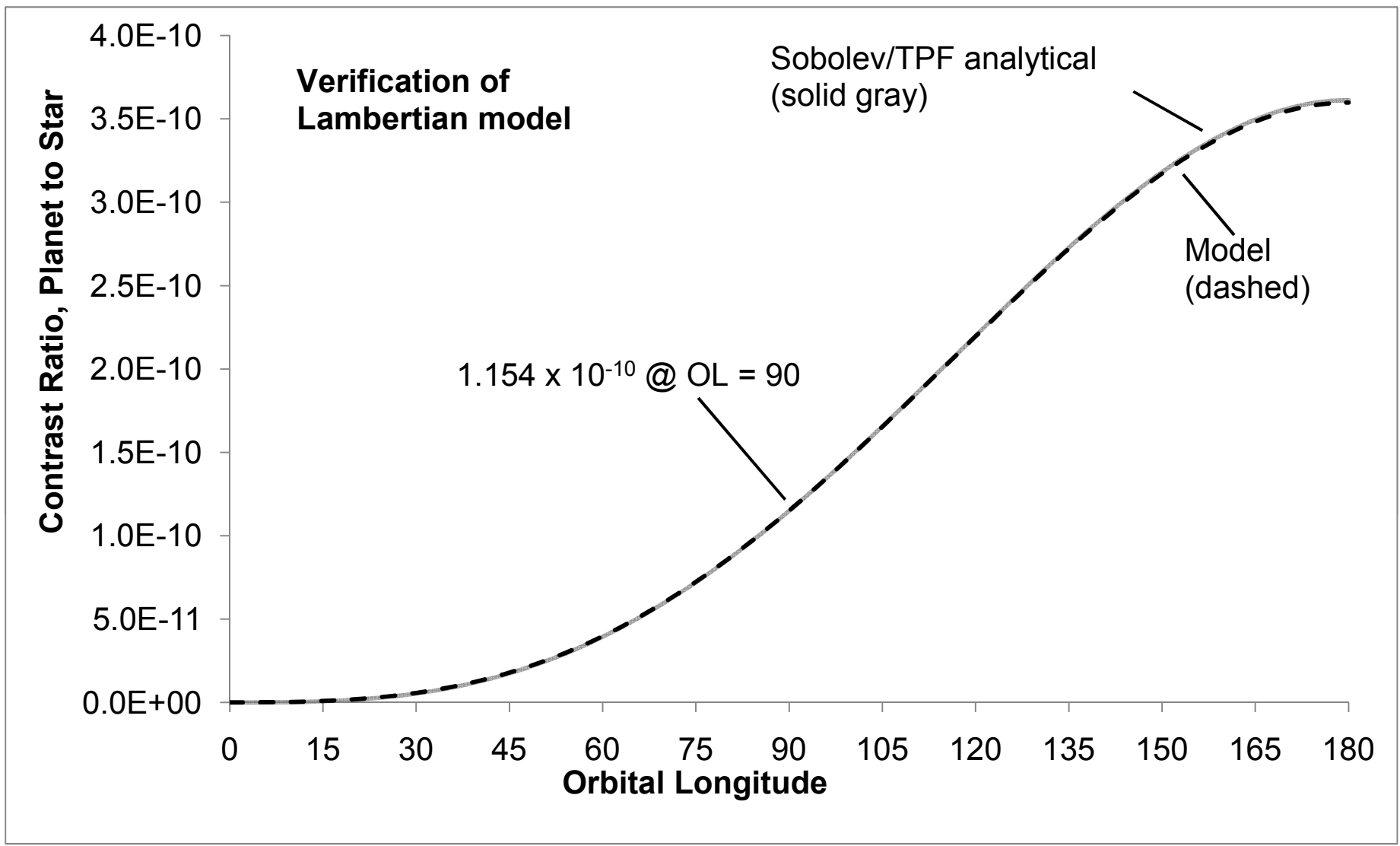

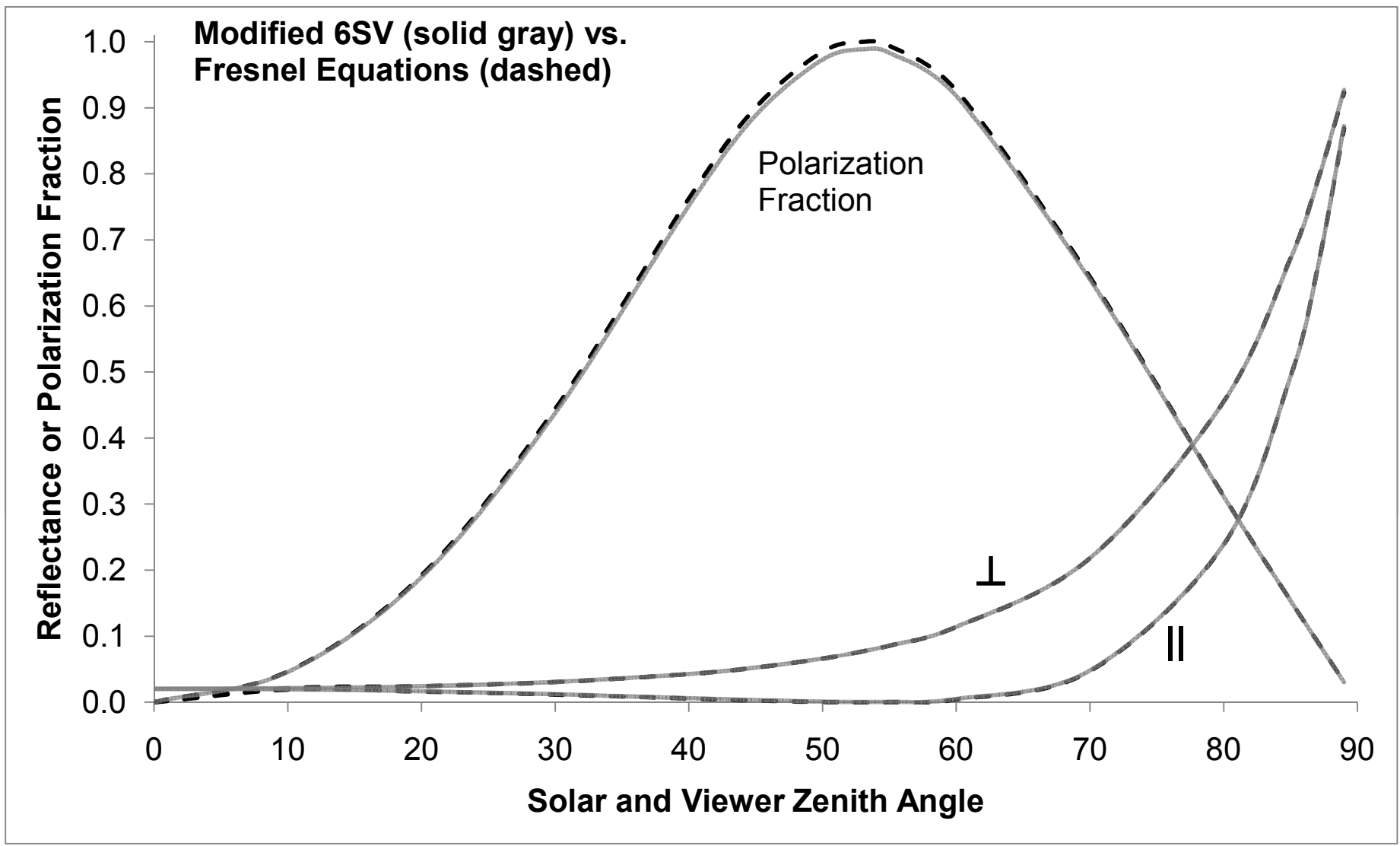

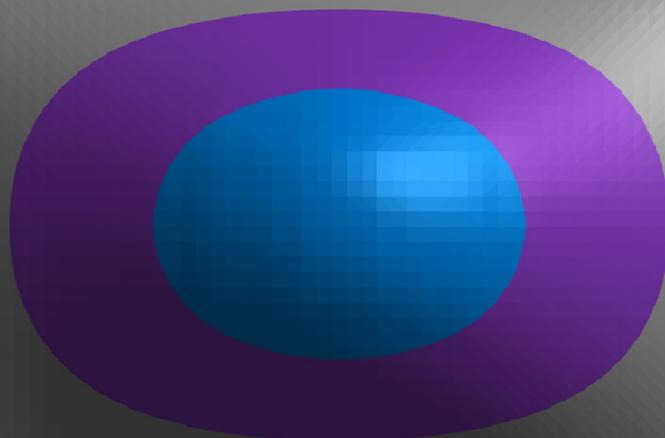
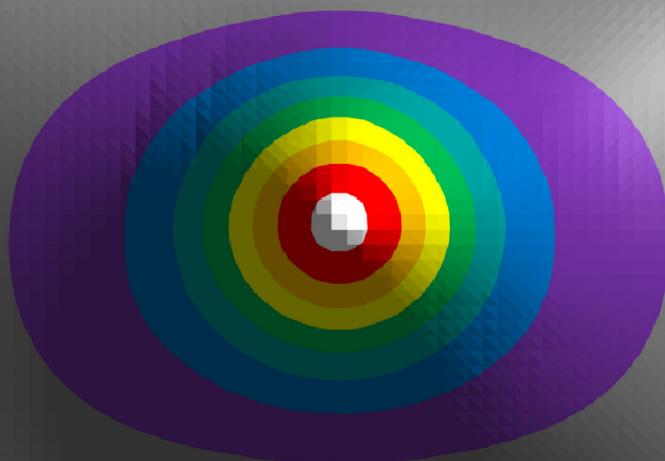
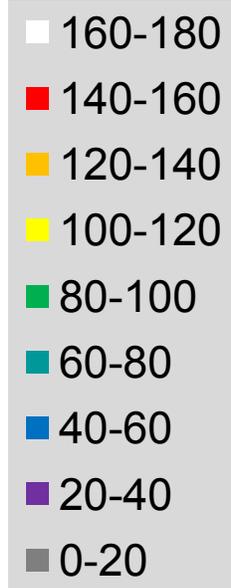

a. parallel

b. perp.

Glint spot, OL = 74